\documentclass[preprint,onecolumn,aps]{revtex4}
\usepackage{amssymb}
\usepackage{amsmath}
\usepackage{graphicx}
\usepackage{subfigure}

\begin{document}

\title{Andreev reflection enhanced single hole tunneling in Ge/Si
  core/shell nanowire quantum dot}

\author{Xiao-Jie Hao$^{(1,2)}$}
\author{Guo-Ping Guo$^{(1)}$}
\author{Hai-Ou Li$^{(1)}$}
\author{Cheng Zhou$^{(1)}$}
\author{Gang Cao$^{(1)}$}
\author{Guang-Can Guo$^{(1)}$}

\author{Wayne Y. Fung$^{(2)}$}
\author{Zhongqing Ji$^{(2)}$}
\author{Wei Lu$^{(2)}$}
\email{wluee@eecs.umich.edu}

\author{Tao Tu$^{(1)}$}
\email{tutao@ustc.edu.cn}

\affiliation{$^{(1)}$Key Laboratory of Quantum Information, University
  of Science and Technology of China, Chinese Academy of Sciences,
  Hefei 230026, People's Republic of China\\
  $^{(2)}$Department of Electrical Engineering and Computer Science,
  The University of Michigan, Ann Arbor, Michigan 48109, USA}
\date{\today }

\begin{abstract}
  We experimentally study the electrical transport properties of Ge/Si
  core/shell nanowire device with two superconducting leads in the
  Coulomb blockade regime. Anomalous zero field magneto-conductance
  peaks are observed for the first time at the gate voltages where
  Coulomb blockade oscillation peaks present. Many evidences indicate
  this feature is due to Andreev reflection enhanced phase coherent
  single hole tunneling through the quantum dot, which can be
  suppressed by an external magnetic field without destroying the
  superconducting states in the electrodes.
\end{abstract}

\maketitle


The novel and fruitful electrical transport phenomena when carbon
nanotubes \cite{Kasumov1999S, Morpurgo1999S, Buitelaar2002PRL,
  Jarillo-Herrero2006N, Jorgensen2006PRL, Cleuziou2007PRL},
semiconductor nanowires \cite{Doh2005S, Dam2006N, Xiang2006NN,
  Sand-Jespersen2007PRL, Doh2008NL, Li2010} or graphene
\cite{Heersche2007N, Miao2007S, Shailos2007EL, Du2008PRB,
  Ojeda-Aristizabal2009PRB, Kessler2010PRL} are connected to
superconductor have attracted special attention in recent years. In
these nano-scale devices, transport properties will highly depend on
the transparency of the interface between the superconducting
electrodes and the nano-structure embedded between them. For fully
transparent contacts, proximity effect will induce supercurrent and
multiple Andreev reflections in the device \cite{Kasumov1999S,
  Doh2005S, Dam2006N, Xiang2006NN, Heersche2007N, Miao2007S,
  Du2008PRB, Ojeda-Aristizabal2009PRB, Shailos2007EL,
  Cleuziou2007PRL}. For intermediately transparent contacts,
interaction between Andreev reflection and Fabry-Perot interference
\cite{Jarillo-Herrero2006N,Jorgensen2006PRL} or Kondo resonance
\cite{Buitelaar2002PRL, Sand-Jespersen2007PRL} can be observed. For
lowest transparent contacts, the non-superconductor between two
electrodes forms a quantum dot, and only superconducting gap opened in
Coulomb blockade regime presents \cite{Doh2008NL, Li2010}.
Interestingly, we find that our experimental system just falls into
the region between the last two situations mentioned above, which
shows Andreev reflection enhanced single hole tunneling (SHT) through
the quantum dot at the center of Coulomb blockade oscillation peaks,
instead of SHT blocked by superconducting gap in electrodes. Moreover,
in the previous investigations of similar structures,
magneto-conductance through the sample was typically not studied.
Here, we measured the magneto-conductance and observed that the phase
information during the Andreev reflection enhanced SHT process can be
destroyed by an external magnetic field before the superconducting
states quenching in the electrodes. Such a SHT tunneling mechanism has
not been observed and reported previously.

\begin{figure}[tbp]
  \centerline{\subfigure[Coulomb blockade]
    {\label{fig:GVg}\includegraphics[width=0.45\columnwidth]{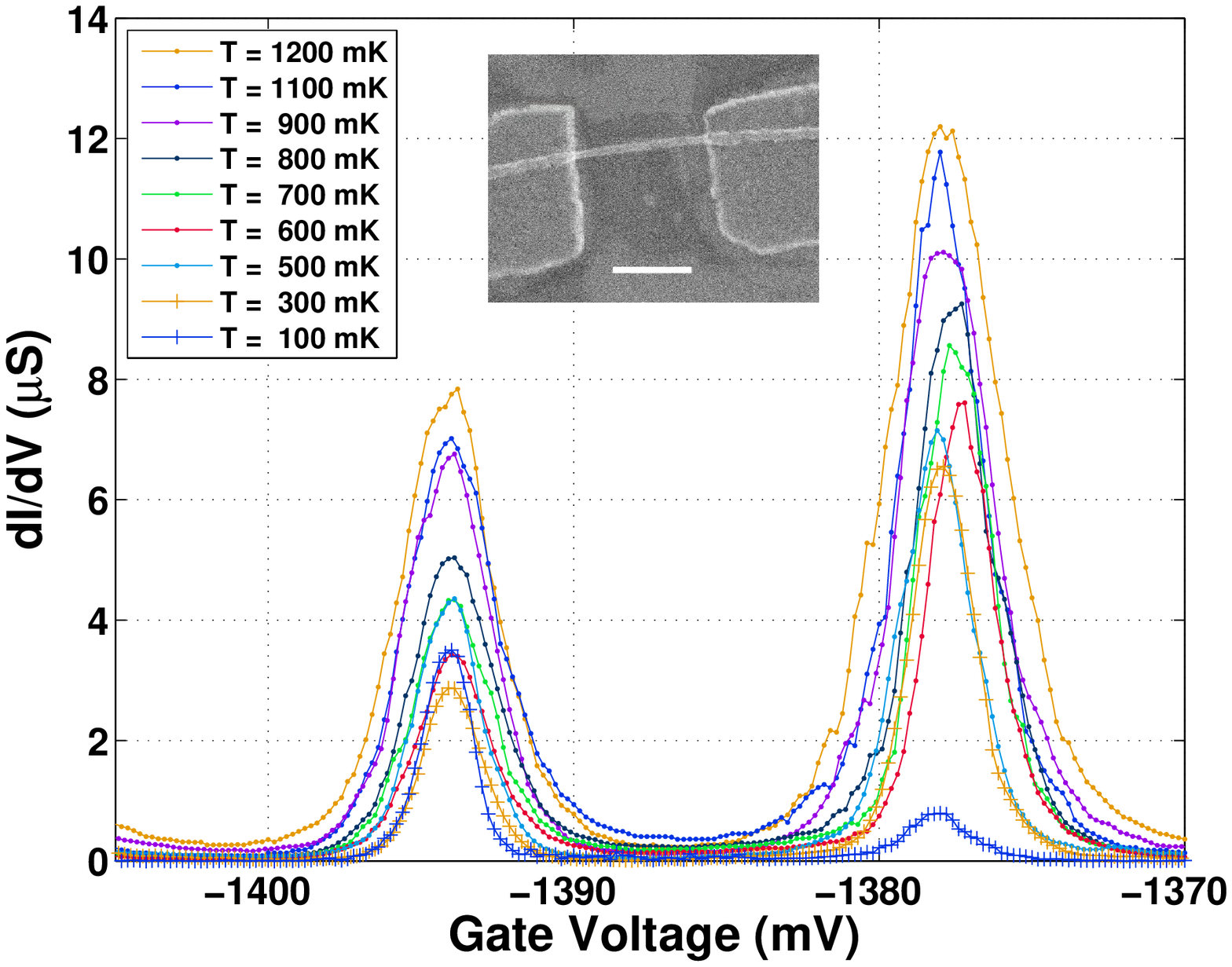}}
    \subfigure[Conductance peak]
    {\label{fig:GB}\includegraphics[width=0.45\columnwidth]{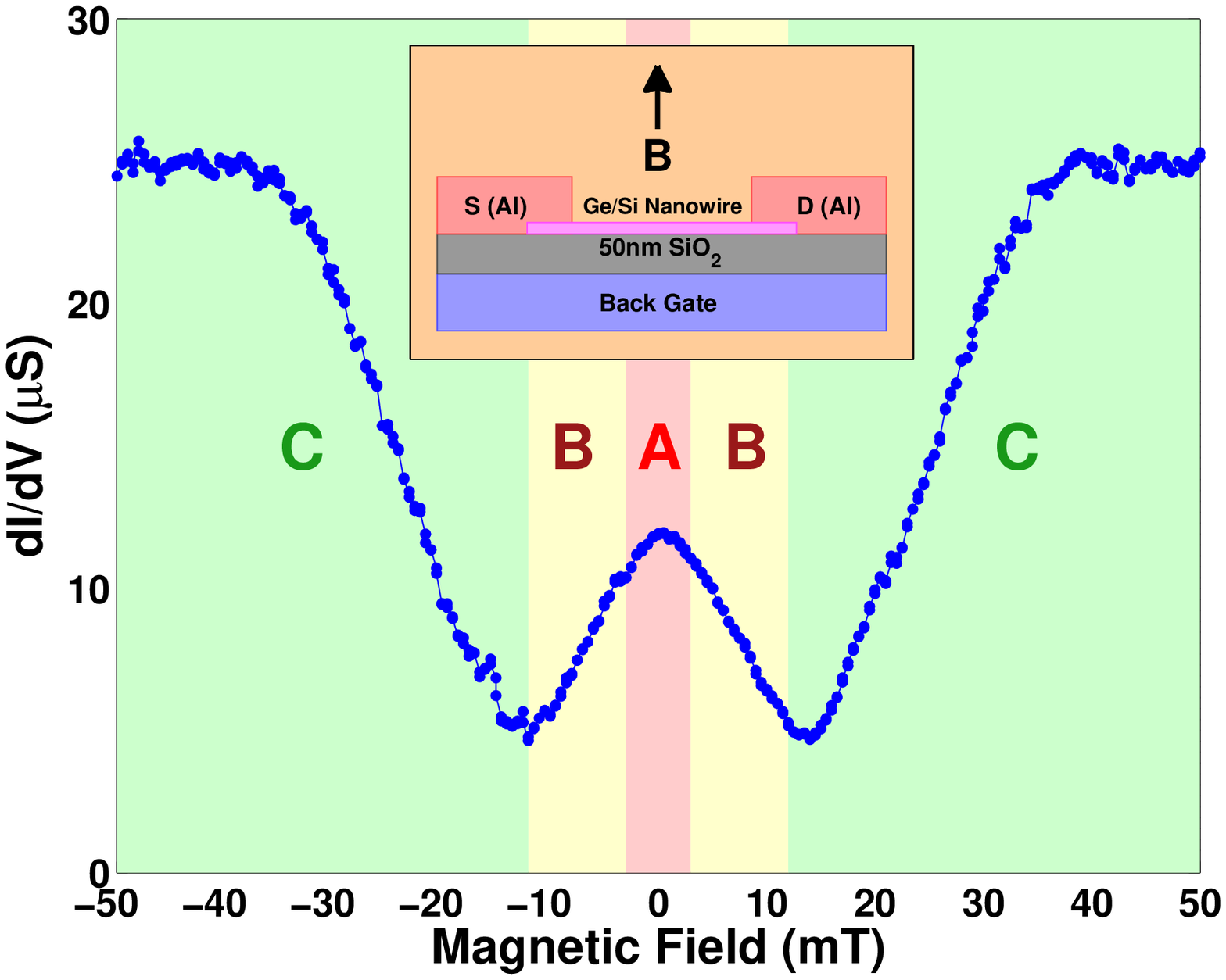}}}
  \caption{~\subref{fig:GVg} Temperature dependence of the
    differential conductance through the device as a function of back
    gate voltage (Coulomb blockade oscillations) at magnetic field
    $B=0 mT$. (inset) SEM image of one device (scale bar: $200 nm$).
    ~\subref{fig:GB} Plot of differential conductance as a function of
    magnetic field, labeled by three different regions `A', `B' and
    `C' ($T=50 mK$ , $V_{SD(AC)}=4 \mu V$). (inset) Experiment setup.}
  \label{fig:GVg_GB}
\end{figure}

The devices reported were fabricated on undoped Ge/Si core/shell
nanowires grown by two-step chemical vapor deposition
\cite{Lu2005PNASUSA}. Due to the large valence band offset between the
Ge core and the Si shell, one-dimensional hole gas can be confined in
the Ge channel. Then we wet transferred nanowires to heavily doped Si
wafer with $50 nm$ silicon oxide, which was used as a back gate in the
measurement. After locating the nanowires relative to the predefined
markers on substrate using Scanning Electron Microscope (SEM), source
and drain contacts were designed by electron-beam lithography. To
remove the native oxide outside the Si shell of the nanowires for good
contact, we immersed the sample in buffered hydrofluoric acid for $3$
seconds. Then $40 nm$ thickness of superconducting Aluminum (Al) were
deposited. Fortunately, without annealing, we still obtained nearly
ideal Ohmic contacts between the Al leads and the nanowires
\cite{Li2010}. SEM image of one of our samples is shown in the inset
of Fig.~\ref{fig:GVg}.

\begin{figure}[thbp]
\centering{\includegraphics[width=0.65\columnwidth]{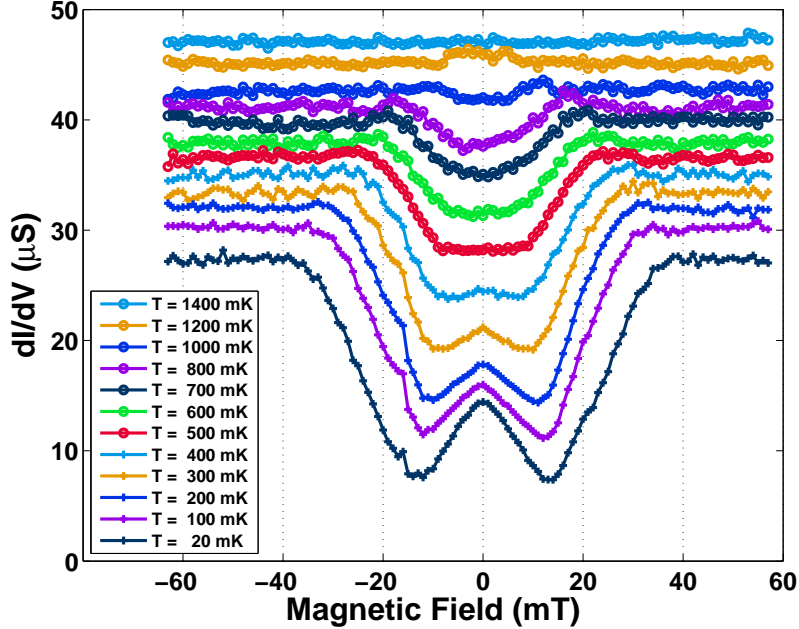}}
\caption{Plot of magnetic field dependent differential conductances at
  various of temperatures. The curves from top to down are taken at $T
  = 1400$, $1200$, $1000$, $800$, $700$, $600$, $500$, $400$, $300$,
  $200$, $100$, $20 mK$, and are vertically shifted for clarity. The
  Andreev reflection enhanced SHT peak vanishes above $500mK$.}
\label{fig:GBT}
\end{figure}

The measurements were performed in a top-loading dilution refrigerator
with an environment base temperature of $20 mK$. In the measurements,
We employed the standard AC lock--in technique with an excitation
voltage of $4\mu V$ at $11.3 Hz$. Similar behaviors were observed in
several devices, and here we show data obtained from two of them
(Fig.~\ref{fig:GVg_GB} and Fig.~\ref{fig:GBT} are taken from one
sample, and the others are from the other one). Using the back gate
$V_{g}$ to tune the energy levels in the quantum dot, we saw clear
Coulomb blockade peaks at a wide range of $V_{g}$. From the Coulomb
blockade oscillations, we obtained an average gate voltage separation
of $\Delta V_{g} \approx 15 mV$, as shown in Fig.~\ref{fig:GVg}. Using
the cylinder-on-plane model \cite{Lu2005PNASUSA}, we calculated the
effective length of the quantum dot to be around $150 nm$, which is
consistent with the sample size. After considering the pinch off gate
voltage of around $V_{g}=2 \sim 3 V$, we estimated $200\sim300$ holes
are typically left in the dot.

We then measured the magneto-conductance by fixing $V_{g}$ at one of
the Coulomb blockade oscillation peaks and applying a magnetic field
perpendicular to both the axis of the nanowire and the substrate. For
very low interface transparency sample, SHT will be blocked when the
energy difference between the source and the drain is less than the
quasi-particle superconducting gap $2\Delta_{Al} \approx 300 \mu eV$
in the Al-leads \cite{Krstic2003PRB}. So in the stability diagram, in
which differential conductance $dI/dV$ is plotted as a function of
source drain bias $V_{SD(DC)}$ and gate voltage $V_{g}$, a gap of
$4\Delta_{Al}/e$ ($e$ is the element charge) will be opened at the
cross point of two adjacent Coulomb diamonds \cite{Doh2008NL}.
Meanwhile, a zero conductance dip with width of $2B_{c}$ at magnetic
field $B=0mT$ \cite{Li2010} will be observed in the
magneto-conductance data. Here $B_{c} \approx 10mT$ is the maximal
magnetic field to preserve Al electrodes in superconducting states,
which is so called critical magnetic field. This kind of dip can still
be seen in our data as shown in Fig.~\ref{fig:GB}. But different from
previous experiments, we find a remarkable peak at the center of the
expected differential conductance dip region. From the conductance
values at Coulomb blockade peaks in normal state ($\left\vert
  B\right\vert > 40 mT$ parts in Fig.~\ref{fig:GB}), we estimated the
tunneling rate between the source/drain lead and the quantum dot is
around $2.5\mu eV$. This value is still smaller than the
superconducting gap $2\Delta_{Al}$, but much larger than the values,
which are around $0.1\mu eV$, we obtained from the low transparency
samples \cite{Li2010}.

\begin{figure}[htbp]
  \centering{\subfigure[$B=0 mT$]
    {\label{fig:GVVg_0mT}\includegraphics[width=0.3\columnwidth]{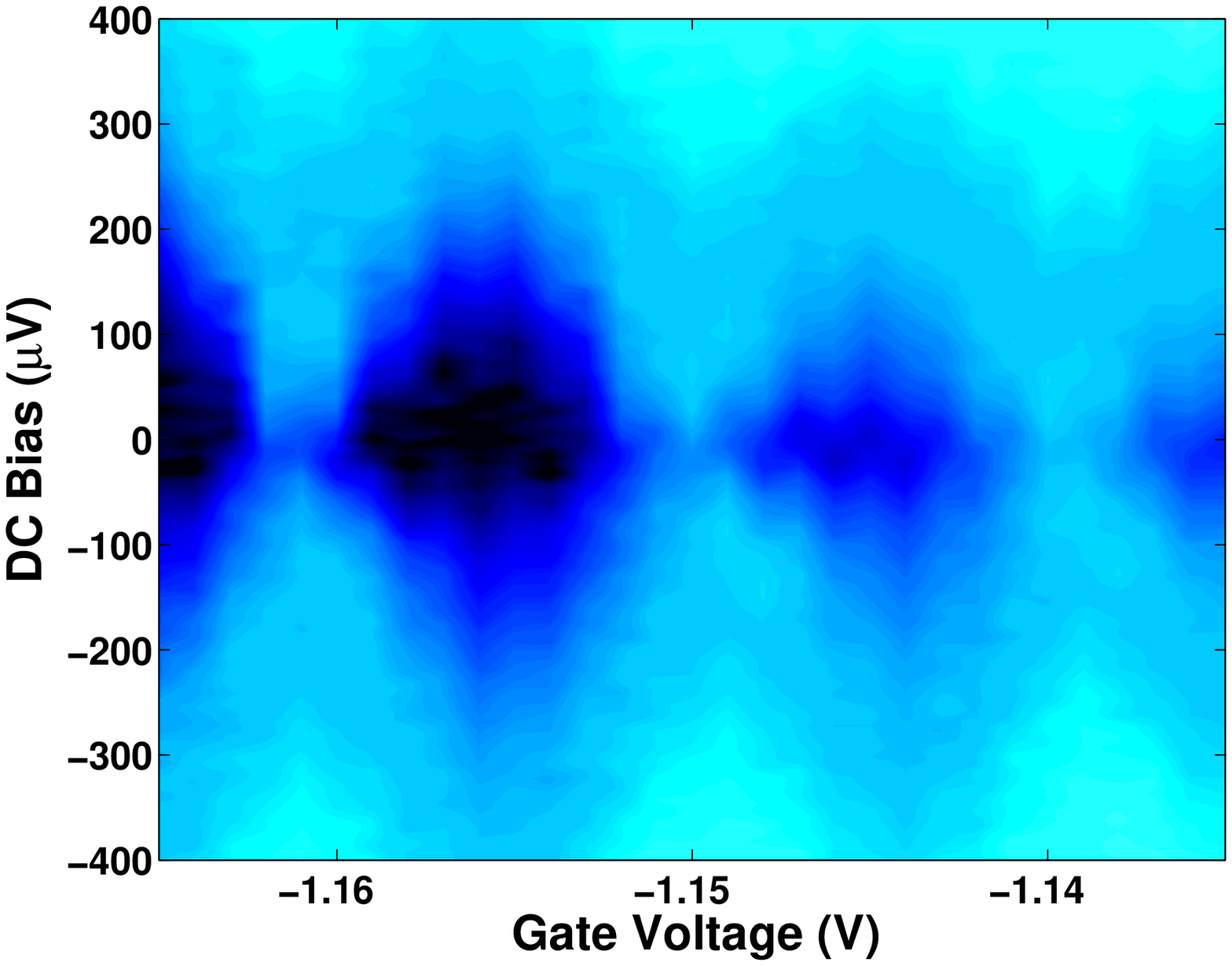}}
    \subfigure[$B=12 mT$]
    {\label{fig:GVVg_12mT}\includegraphics[width=0.3\columnwidth]{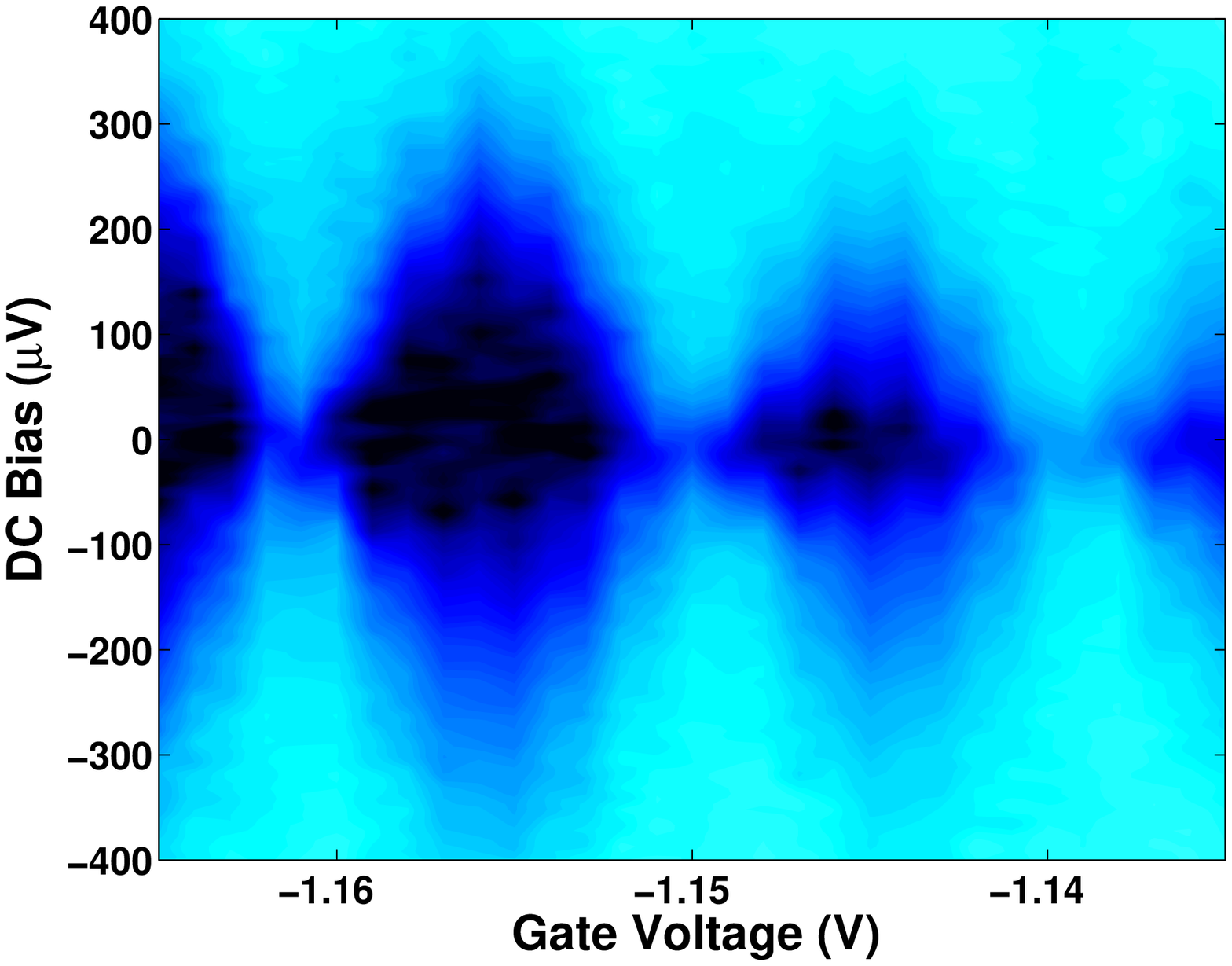}}
    \subfigure[$B=60 mT$]
    {\label{fig:GVVg_60mT}\includegraphics[width=0.3\columnwidth]{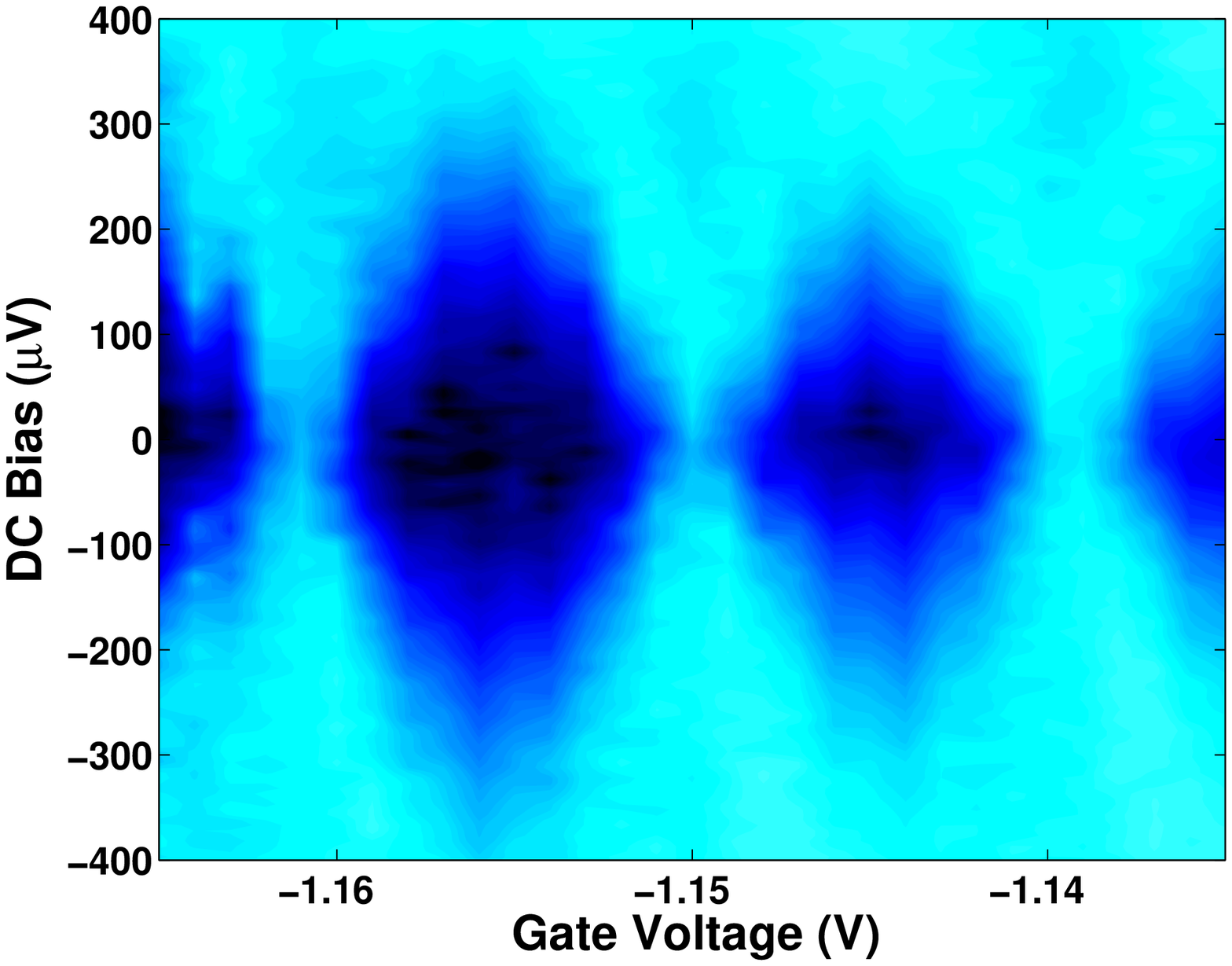}}}
  \caption{Stability diagram (differential conductance (unit: $\mu S$)
    as a function of DC bias voltage and back gate voltage) at
    ~\subref{fig:GVVg_0mT} $B=0mT$, ~\subref{fig:GVVg_12mT} $B=12mT$
    and ~\subref{fig:GVVg_60mT} $B=60mT$ with same color axis ($T=50
    mK$, $V_{SD(AC)}=4 \mu V$). In the light blue (dark black) parts,
    differential conductance $dI/dV$ has large (small) values. The gap
    at the center of each diamond opens more from $B=0mT$ to $B=12mT$
    then gradually closes when $B$ approaching $60mT$.}
  \label{fig:GVVg}
\end{figure}

In order to illustrate the evolution of the stability diagram during
the magnetic field increasing, we plotted three stability diagrams in
Fig.~\ref{fig:GVVg}, at different magnetic fields $B=0mT$
(Fig.~\ref{fig:GVVg_0mT}), $B=12mT$ (Fig.~\ref{fig:GVVg_12mT}), and
$B=60mT$ (Fig.~\ref{fig:GVVg_60mT}). The light blue (dark black) parts
in these figures represent large (small) values of differential
conductance $dI/dV$. It is obvious that the gap at $V_{SD(DC)}=0$,
which should be widest at $B=0mT$ according to previous reports
\cite{Doh2008NL,Li2010}, is indeed a little bit closed at $B=0mT$ in
our system. The gap starts to open as the magnetic field growing up,
and reaches a maximum at $B=12mT$ but is still significantly less than
$4\Delta_{Al}/e$. After $B>B_{c}$, this gap gradually vanishes along
with the Al leads turning into normal (non-superconducting) states.

\begin{figure}[htbp]
\centering{\includegraphics[width=0.65\columnwidth]{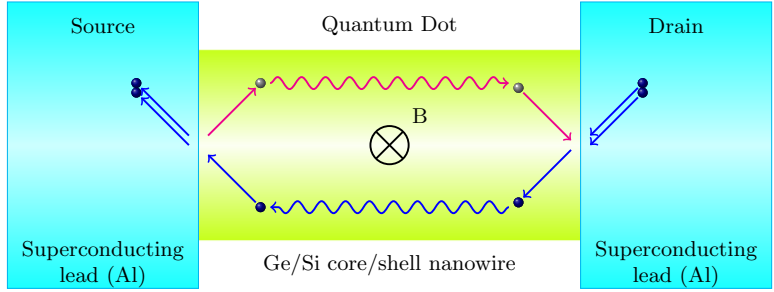}}
\caption{Schematic diagram of the Andreev reflection enhanced phase
  coherent SHT through the quantum dot. An external magnetic field
  perpendicular to the nanowire will destroy the phase information in
  the transport loop caused by Andreev reflections. The reflected
  particles should have the opposite momentum of incident particles,
  but here we plot them in different directions for clarity.}
\label{fig:AR}
\end{figure}

The abnormal magneto-conductance presented here is apparently
different from the ones reported before \cite{Li2010}, which implies
that a new transport mechanism exists in this system. First of all, we
did not observe any evidence of Kondo effect in any of our devices
having these magneto-conductance features. In addition, the Kondo
resonance induced effect occurs in the dips of Coulomb blockade
oscillations \cite{Buitelaar2002PRL}, while the anomalous peak here
were observed at the peaks of Coulomb blockade oscillations. These
arguments exclude the influence of Kondo resonances to the SHT
process. Since the magneto-transport shows negative
magneto-conductance at very low field, weak antilocalization could be
one possible origination of the zero field peak. However, based on our
previous investigation of weak antilocalization and spin-orbit
coupling on this system \cite{Hao2010WAL}, the drop of conductance
could not be so strong for a change of several milliteslas in magnetic
field. Besides, the high temperature ($T$ larger than the critical
temperature of Al: $T_{c} \approx 1.2K$) data does not show this kind
of peak, but the weak antilocalization survives even as the
temperature goes beyond $100K$. Here, we explain our data as Andreev
reflection enhanced SHT through the quantum dot. In the Andreev
reflection, when an electron (hole) from non-superconducting material
incidents at the interface of superconductor and non-superconductor,
it will be reflected with the opposite spin and momentum as a
phase-conjugated hole (electron), and forms a Cooper pair in the
superconductor \cite{Andreev1964SPJ}. In our system, as shown in
Fig.~\ref{fig:AR}, one hole runs in the nanowire and incident at the
interface of the nanowire and the drain, it will be reflected as an
electron in the nanowire and also destroys a Cooper pair in the drain.
Then the electron travels through the nanowire and incidents at
another interface, the one between the nanowire and the source, it
will be reflected as a hole and create a Cooper pair in the source.
This theory well explains the features observed in Fig.~\ref{fig:GB}.
There are three regions labeled as `A', `B' and `C'. In the region
`A', even if there are superconducting gaps in the source and drain
leads, current can still flow through the device continuously at zero
bias. In this way, the SHT process is actually enhanced by Andreev
reflections. In addition, because of the Andreev reflections can
preserve the phase information, the Andreev reflection enhanced SHT
transport is phase coherent \cite{Doh2005S}. While in region `B', an
external magnetic field less than $B_{c}$ applied perpendicular to the
transport channel will break the phase information during the
transport. That is why the phase coherent transport is suppressed and
the differential conductance decreases while the magnetic field
increases. In region `C', larger magnetic field exceeding the critical
field $B_{c}$ destroys the superconducting states in electrodes. So
the transport will come back to the normal Coulomb blockade
oscillation regime and the conductance will recover to the normal
value.

\begin{figure}[htbp]
  \centering{\subfigure[$dI/dV$ v.s. $B$ \& $V_{SD(DC)}$]
    {\label{fig:GVB}\includegraphics[width=0.45\columnwidth]{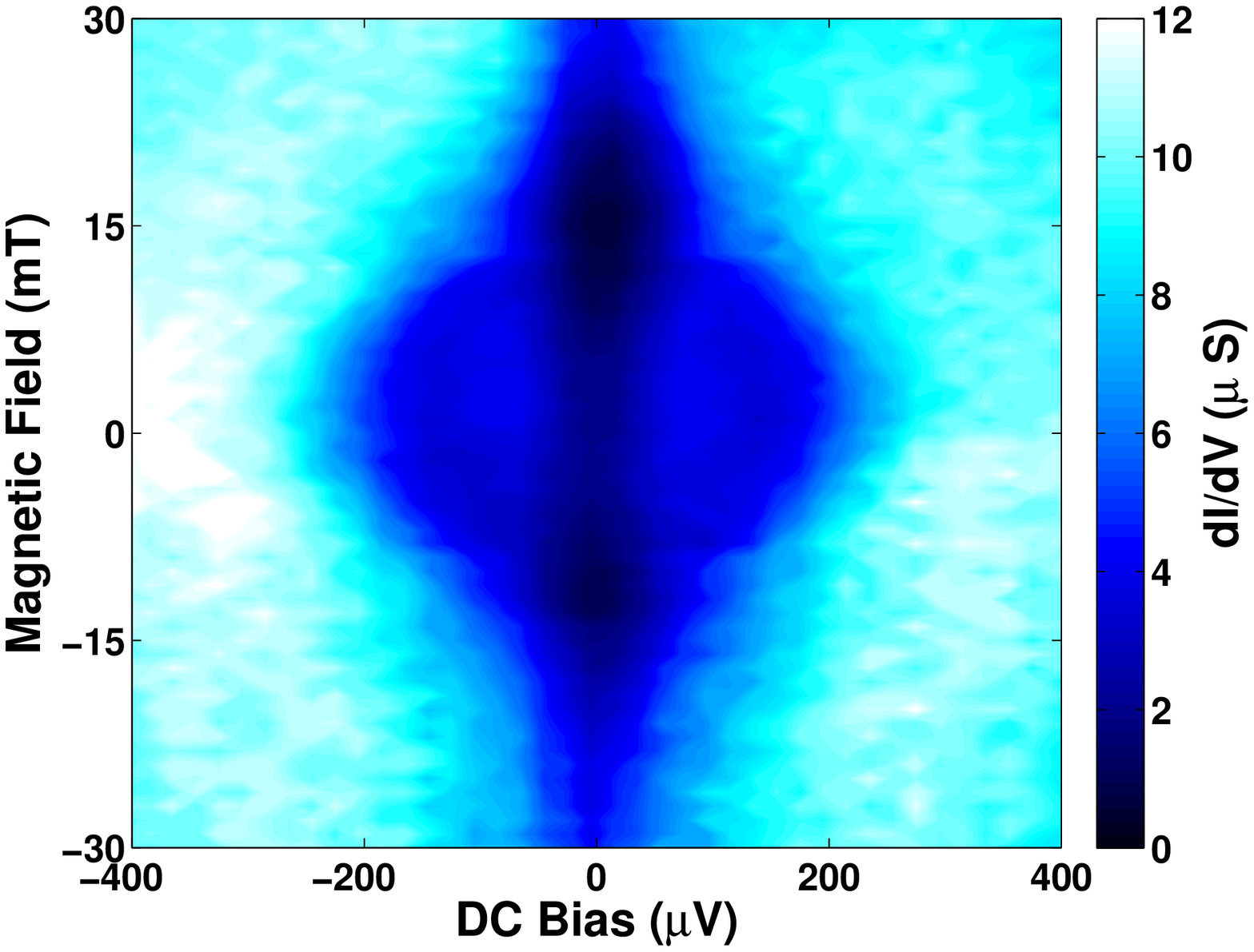}}
    \subfigure[$dI/dV$ v.s. $B$ \& $V_{g}$]
    {\label{fig:GBVg}\includegraphics[width=0.45\columnwidth]{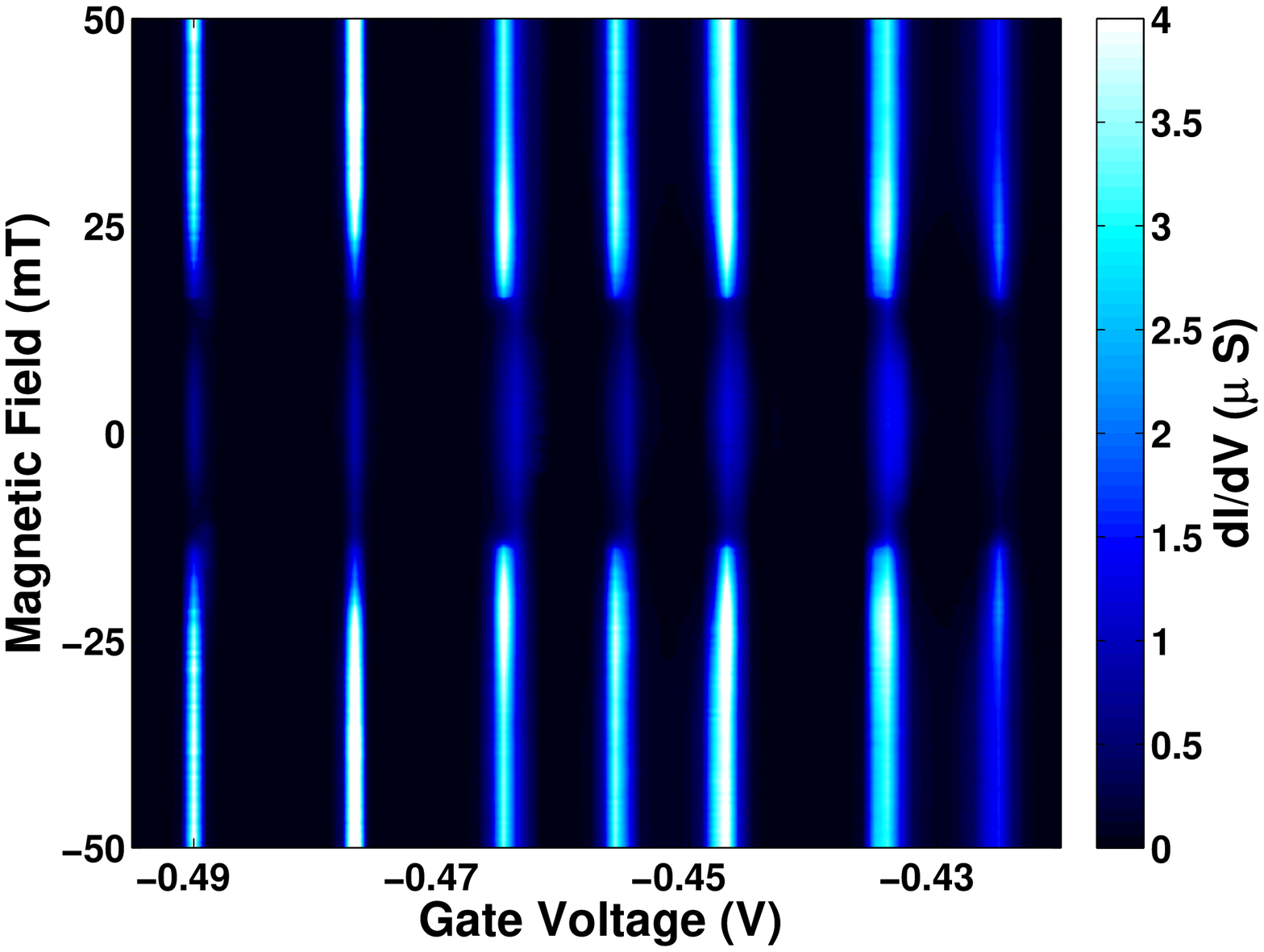}}}
  \caption{Color-scale plots of differential conductance (unit: $\mu
    S$) as a function of ~\subref{fig:GVB} magnetic field and DC bias
    voltage and ~\subref{fig:GBVg} magnetic field and gate voltage.}
  \label{fig:GVB_GBVg}
\end{figure}

The Andreev reflection enhanced SHT explanation was supported by
additional data. In Fig.~\ref{fig:GVB}, the differential conductance
$dI/dV$ is plotted against the magnetic field $B$ and the DC bias
voltage $V_{SD(DC)}$ while fixing $V_{g}$ at the Coulomb blockade
oscillation peak. In this figure, it is clearly seen that at
$\left\vert V_{SD(DC)}\right\vert<2\Delta_{Al}/e$, $dI/dV$ has a local
maximum at zero field and decreases as $B$ increasing, then goes up
and stabilizes after $\left\vert B\right\vert > B_{c}$. The
magneto-conductance feature is observed not only at $V_{SD(DC)}=0$ but
also at all voltages $\left\vert V_{SD(DC)}\right\vert<2\Delta_{Al}/e$
because of inelastic tunneling. The temperature dependence of the
Coulomb blockade oscillations shown in Fig.~\ref{fig:GVg} implies the
energy level splitting in the quantum dot is much smaller than
$K_{B}T$, where $K_{B}$ is Boltzmann's constant and $T$ is the
temperature of charge carriers. It means the dot we measured was in
classical Coulomb blockade regime, where inelastic tunneling can
easily happen. This is consistent with the quantum dot being in the
many--hole regime, confirmed by the holes left in the dot discussed
before. In addition, as shown in the light blue regions in
Fig.~\ref{fig:GVVg}, differential conductance keeps increasing also
indicates inelastic tunneling through the quantum dot. Therefore, in
our quantum dot, as a result of excited states falling into the gap
broadened by temperature and inelastic tunneling, Andreev reflection
enhanced SHT can still happen even if the ground state energy level in
the quantum dot is not exactly aligned with the Fermi levels of the
source or drain.

Gate voltage dependence of magneto-conductance is displayed in the
color-scale plot of Fig.~\ref{fig:GBVg}, in which we find the
interesting magneto-conductance peak feature appears at each Coulomb
oscillations peak. But the strength of the peak depends on the Coulomb
oscillation in the normal state. This is consistent with that the
probability of Andreev reflection occurs depends on the effective
transparency of the interface, i.e. the alignment of energy levels in
the quantum dot with Fermi levels in the source and drain electrodes
\cite{Buitelaar2003PRL, Jarillo-Herrero2006N}. By means of a gate
voltage, the Andreev enhanced SHT are modulated as well as the Coulomb
oscillations. This gate tunable phenomenon is similar as reported by
Jarillo-Herrero \textit{et al.} \cite{Jarillo-Herrero2006N}. In their
experiments, the supercurrent and multiple Andreev reflections can be
turned on and off by gating the carbon nanotube at different
Febry-Perot interference regions. Notably, the normal state
conductance in the region where supercurrent observed in their device
is slightly higher than ours but in the same order of $e^{2}/h$ ($h$
is the Planck's constant). The key factor which is responsible for the
Andreev reflection enhanced phase coherent SHT in our system is the
interface transparency. The interface in our case is a little opaque
but not opaque enough to prevent the Andreev reflection process.
Therefore, in our case the transport is dominated by SHT events and
the gate voltage tunes the position of the energy levels and opens a
channel at the position where SHT is enhanced through multiple Andreev
reflections.

We also tested the device at different temperatures. From the data
given in Fig.~\ref{fig:GBT}, we find that the conductance dip owing to
SHT blocked by superconducting gap collapses and finally disappears at
$T>T_{c}$ \cite{Li2010}. But the Andreev reflection enhanced SHT
feature vanishes before Al leads lose superconductivity at around
$T=500mK$, at which temperature the superconducting gap in Al starts
to decrease dramatically according to the superconductivity theory
\cite{Li2010}, and does not show up after $T>T_{c}$. These
experimental phenomena are consistent with the explanation in
Fig.~\ref{fig:AR}, and suggest that the phase coherent transport is
directly correlated to the superconducting states in the electrodes
and is much easier to be destroyed by heating effect than the
superconducting state in Al.

In conclusion, we observed magneto-conductance peak at zero magnetic
field on superconductor contacted Ge/Si core/shell nanowire quantum
dot at the point of Coulomb blockade oscillation peak. The
experimental data are consistent with our explanation --- Andreev
reflection enhanced phase coherent single hole tunneling through the
quantum dot. Neither the demonstrated results nor suggested transport
mechanism has been reported before in previous literature, and both of
them justify further exploring experimentally and theoretically.

This work was supported by the National Basic Research Program of
China (Grants No. 2009CB929600, No. 2006CB921900), the National
Natural Science Foundation of China (Grants No. 10804104, No.
10874163, No. 10604052, 10934006), and the US National Science
Foundation (ECS-0601478).

\bibliography{AR_SHT}

\end{document}